\documentstyle{elsart}
\newcommand{\tres}[1]{\buildrel\ldots\over #1}
\begin{document}
\begin{frontmatter}

\title{Generalized power expansions in cosmology}

\author{Alejandro S. Jakubi}
\address{Departamento de F\'{\i}sica,
Facultad de Ciencias Exactas y Naturales,
Universidad de Buenos Aires,
Ciudad  Universitaria,  Pabell\'{o}n  I,
1428 Buenos Aires, Argentina.}

\begin{abstract}
It is given an algorithm to obtain generalized power asymptotic expansions of
the solutions of the Einstein equations arising for several homogeneous
cosmological models. This allows to investigate their behavior near the
initial singularity or for large times. An implementation of this algorithm in
the CAS system Maple V Release 4 is described and detailed calculations for
three equations are shown.
\end{abstract}
\end{frontmatter}

\section{Introduction}

For several homogeneous cosmological models the Einstein equations, combined
with the constitutive equations for the matter source, can be cast into
nonlinear ordinary differential equations of the form

\begin{equation} \label{dy}
D[y(t)]=\sum_{i=1}^{N} A_i\, y^{B_i^0}\left(\frac{dy}{dt}\right)^{B_i^1}
\cdots \left(\frac{d^r y}{dt^r}\right)^{B_i^r}=0
\end{equation}

\noindent where $y$ is usually a monotonic function of either the scale factor
or the Hubble rate, $t$ is the universal time, $A_i$ and $B_i^p$ are real
constants, $r$ is the order of the ODE and without loosing generality we can
assume that $B_i^p\ge 0$ by eventually multiplying the equation by suitable
powers of $y$ and its derivatives \cite{Dav77} \cite{Chi89} \cite{Pav}
\cite{AFI}. Equation (\ref{dy}) is well defined provided we restrict to
$y(t)\ge 0$. This restriction appears naturally in the cosmological setting
because the scale factor is positive, or the Hubble rate is positive for an
expanding universe.

When an exact solution of (\ref{dy}) is not available, one is at least
interested in obtaining some information about it in the form of an expansion
for the limits $t\to 0^{+}$ (usually corresponding to the behavior of the
solution near the initial singularity) and $t\to\infty$. However solutions to
equations of the form (\ref{dy}) frequently do not have power series solutions
with integer or even rational exponents (Pusieux series) in these limits. Thus
we are led to try "generalized" power series expansions of the form

\begin{equation} \label{yt}
y(t)\sim \sum_{j=1}^\infty c_j t^{n_j}
\end{equation}

\noindent where $c_j$ and $n_j$ are real constants and $n_1<n_2<\cdots$ for
$t\to 0^+$ and $n_1>n_2>\cdots$ for $t\to \infty$. So $c_1 t^{n_1}$ is the
leading term, $t^{n_{i+1}}/t^{n_i}\to 0$ in either limit and the set
$\left\{t^{n_i}\right\}$ constitutes an asymptotic scale. Inserting this
expansion in (\ref{dy}) and performing the necessary asymptotic expansions we
get

\begin{equation} \label{dsum}
D\left[\sum_{j=1}^\infty c_j t^{n_j}\right]\sim \sum_{k=1}^\infty C_k t^{e_k}
\end{equation}

\noindent where the exponents $e_k$ are real and form an ordered set:
$e_1<e_2<\cdots$ for $t\to 0^+$ and $e_1>e_2>\cdots$ for $t\to \infty$. So, if
the equation (\ref{dy}) admits a solution with expansion (\ref{yt}), each of
the $C_k$ must vanish and this set of equations fix in principle the constants
$c_j, n_j$ in (\ref{yt}) up to $r-1$ free parameters arising from the
integration constants of the general solution of (\ref{dy}) (for simplicity in
the notation  the arbitrary constant corresponding to time
translation freedom is fixed to $0$).

Several critical steps in the search for solutions in the form of generalized
power expansions involve calculations with large number of terms and this
number grows very fast with the size of the ODE making hand calculation
inconvenient. However, as many of steps in these calculations have a
mechanical nature, use of computer algebra systems appear ideally suited.

The algorithm of used in obtaining these approximated solutions has partial
similarity in its initial steps to the algorithm used to test whether a given
ODE satisfies necessary conditions to has Painlev\'e property, namely that its
solutions are single valued around movable singularities \cite{ARS}. That algorithm
checks whether the behavior of the solutions near the singularities is like
(\ref{yt}) with integer exponents. However in this paper the study of
solutions is restricted to the positive real axis and  no consideration will
be made about their multivaluedness when extended to the complex plane.

\section{The algorithm}

The objective of calculations is to obtain a truncation of (\ref{yt}) to
a finite number of terms, say $M$

\begin{equation} \label{yM}
y_M(t)= \sum_{j=1}^M c_j t^{n_j}
\end{equation}

\noindent The method of calculation is iterative, so that constants $c_M,
n_M$, $M>1$, when not free are determined by $c_1,n_1,\ldots,c_{M-1},n_{M-1}$.
For each step of the iteration in $M$, when inserted $y_M$ in (\ref{dy}), two
critical points of the calculation are:

(i) Asymptotic expansion (to order $M$) of the
terms with noninteger $B_i^p$.

(ii) Collect all the terms with the same power of $t$.

\noindent
Thus we arrive at an expression of the form

\begin{equation} \label{dsumM}
D\left[y_M(t)\right]= \sum_{l=1}^R D_l \, t^{f_l}
\end{equation}

\noindent where $D_l$ and $f_l$ are real constants. In general the sequence of
exponents $f_1, f_2, \ldots, f_R$ is not ordered. Thus, for every step the
next task is:

(iii) Sort the $\left\{f_l\right\}_{1\le l\le R}$.

\noindent This sorting operation is perhaps the most involved part of the
whole calculation. To describe its role and a procedure to do it, only the
behavior for $t\to 0^+$ will be considered in the following.

We start with $M=1$, that is $y_1=c_1t^{n_1}$. When it is 
inserted in the i-th term of ( \ref{dy} ) we get a term with exponent\begin{equation} \label{g}
g_i=\sum_{h=0}^r \left(n_1-h\right) B_i^h
\end{equation}

\noindent
and coefficient

\begin{equation} \label{E}
E_i=A_ic_1^{\sum_{h=0}^r B_i^h}\prod_{h=1}^r \left(n_1-h+1\right)^
{\sum_{j=h}^r B_i^j}\equiv \nu_i c_1^{\mu_i}
\end{equation}

\noindent
If now we take $M=2$, crossed terms appear. Let us take any one of them, say
all factors from the leading term except that of derivative $b$. Then we get
an exponent

\begin{equation} \label{g'}
g_i'=g_i+\left(n_2-n_1\right) B_i^b > g_i
\end{equation}

\noindent As $e_1$ is the minimum of the $\left\{g_i\right\}_{1\le i\le N}$,
and say that this minimum occurs for $i\in I$, so $C_1=\sum_{i\in I}E_i$ is a
function of $c_1$ and $n_1$, and we find that only the leading term of
(\ref{yt}) can contribute to the leading term of (\ref{dsum}).

The requirement that the leading term of (\ref{dsum}) is not a constant leads
to $\mu_i>0$ for $i\in I$. If $I$ has only one element the requirement $C_1=0$
implies $c_1=0$. This implies that $I$ must have at least two elements. In
this case it is said that these terms balance. This constraint usually
determines $n_1$. For instance if terms $i$ and $k$ balance and
$\mu_i\neq\mu_k$, we get

\begin{equation} \label{n1}
n_1^{ik}=-\frac{\sum_{h=0}^r h\left(B_i^h-B_k^h\right)}
{\sum_{h=0}^r \left(B_i^h-B_k^h\right)}
\end{equation}

\noindent
For this $n_1$ we get $c_1$ provided

\begin{equation} \label{c1}
c_1^{\mu_i-\mu_k}+\frac{\nu_k}{\nu_i}=0
\end{equation}

\noindent
has real roots. On the other hand, if the terms $i$ and $k$ balance and
$\mu_i=\mu_k$, the additional constraint $\sum_{h=0}^r
h\left ( B_i^h-B_k^h\right )=0$ must hold, $c_1$ remains arbitrary and $n_1$
is
determined by the real roots of the equation $\nu_i+\nu_k=0$.

Each pair of real numbers $(c_1,n_1)$ obtained from this analysis  corresponds
to the leading term of the expansion of a family of solutions. Thus,
subsequent calculations must be done separately.

Both the constants $A_i$ and $B_i^h$ in (\ref{dy}) may contain free parameters
of the model. Thus, considering that some of the $B_i^h$ are free, we have to
ask for the values of the parameters that make these $n_1$ coincide. They
are critical values for the parameters as they make the behavior of the
solutions change.

Repeating the argument leading to (\ref{g'}) order by order in $M$ it is easy
to see that $c_M$ and $n_M$ only appear in $C_j$ for $j\ge M$. This
also shows that the first $M$ terms of (\ref{dsumM}), once sorted, are equal
to the first $M$ terms of expansion (\ref{dsum}). In particular, the first
$M-1$ terms are those already found in step $M-1$ of the iteration. Thus the
main tasks at step $M$ are:

(iv) Find $e_M$, $n_M$ (if possible) and $C_M$.

(v) Solve $C_M=0$ for either $c_M$ or $n_M$.

\noindent
Generalizing the expressions (\ref{g}) and (\ref{g'})
we see that the exponents $f_l$ are linear functions of
$n_1,\ldots,n_M$

\begin{equation} \label{fl}
f_l=\sum_{j=1}^M \alpha_l^j n_j+\sum_{i=1}^N\sum_{h=0}^r \beta_l^{ih} B_i^h
\end{equation}

\noindent
where the $\alpha_l^j$ are in turn linear functions of the $B_i^h$.
To order these exponents we need to know first when they
can be equal. Consider that $f_l=f_k$, then we have the equation

\begin{equation} \label{lk}
\sum_{j=1}^M \left(\alpha_l^j-\alpha_k^j\right) n_j+
\sum_{i=1}^N\sum_{h=0}^r \left(\beta_l^{ih}-\beta_k^{ih}\right) B_i^h=0
\end{equation}

\noindent
This equation sets a constraint between the $\left\{n_j\right\}$ or
equivalently defines an $(M-1)$-dimensional hyperplane in the vector space
$\left(n_1,\ldots,n_M\right)$ provided the vector
$\Delta\alpha_{lk}=\left(\alpha_l^1-\alpha_k^1,\ldots,\alpha_l^M-\alpha_k^M\right)\neq
0$.

Only a sector of the $M$-dimensional vector space is admissible. First of all,
the inequalities $n_1<n_2<\cdots<n_M$ must be satisfied. Besides,
$n_1,\ldots,n_{M-1}$ have been determined in the previous steps of the
iteration. As they are usually functions of the parameters contained in
(\ref{dy}), the variation of these parameters within their range further
restricts the admissible sector of the $M$-dimensional space. Within it, if
the hyperplane exist, $n_M$ becomes a linear function of $n_1,\ldots,n_{M-1}$,
and from all pairs of exponents in $\left\{f_l\right\}_{1\le l\le R}$ the $n_M$
must be found for which both exponents are minimum. In this way we find
$e_M=f_l=f_k$, $n_M$ and $C_M=D_l+D_k$. From $C_M=0$ we obtain $c_M$.
On the other hand, if $\Delta\alpha_{lk}=0$, $c_M$ remains free and $C_M=0$
determines $n_M$.

Within the admissible sector, in each subsector delimited by the hyperplanes,
the $f_l$ can be sorted. The minimum, say $f_k$ yields $e_M$ and provided
$D_k$ has a common factor of the form $c_M^\mu$ and the equation $D_k=0$ can
be solved for a real $n_M$, this case also yields the pair $(c_M, n_M)$
with $c_M$ free. In all cases only real solutions are admissible.

\section{Implementation}

The implementation of the algorithm described in the previous section is made
in Maple V Release 4. At present, this implementation involves use of some
routines written in the Maple V programming language as well as interactive
calculations in the worksheet environment.

For the task of collecting terms with equal power of $t$, the command {\bf
collect} of the current implementation is not suitable as it cannot collect
terms with nonrational exponents. So, the first task of the implementation was
to write some procedures to provide this facility. Namely through a command
named {\bf collect2}, given as input an expression, a variable (or an
expression), one or more
options and a procedure it yields, depending on the options,

(a) a form of the expression collected in generic powers of the variable
(assumed nonnegative throughout),

(b) a list of the exponents,

(c) a list of the powers,

(d) the (collected) coefficients in the form of a table, indexed by the
exponents.

\noindent For options (a) and (d) the procedure given as input is used to
output the coefficients is a useful way (typically using {\bf collect} ). The
package {\bf collect2} has also some useful procedures that can be used
independently like {\bf expand1} (like {\bf expand} but does not expand
exponents), and {\bf varsubs} (like {\bf algsubs} but allows
substitution of subexpressions raised to generic powers).

The rest of the calculations are made for each step of the iteration at the
worksheet level. It is planned to automatize some of them when some more
experience is obtained in solving some failures that show the system. One of
the main difficulties appears in the sorting the list of exponents when a set
of inequalities is imposed. Maple provides the {\bf assume} facility by which
properties about variables, in this case inequalities, are informed to the
system so that its routines can make use of this information and take
decisions when handling these variables. The command {\bf is} is given to
check whether a property is true provided that the necessary properties of the
involved variables hold. However, in some cases, this command fails in
deciding the order of two expressions, though mathematically the calculation
is well defined and a hand calculation can demonstrate the inequality in a few
lines. So we use procedures {\bf ord1} and {\bf ord2},  that use numerical
comparisons, based on values that lay inside the interval of interest. These
and some other auxiliary procedures are currently collected in a package named
{\bf general}.

In the following sections three examples will be shown to give a feel of how
this calculations can be done with Maple.

\section{Example with three terms}

The system of Einstein
equations for homogeneous, isotropic cosmological models with a
variety of matter sources reduce to the ordinary
differential equation

\begin{equation} \label{1}
\ddot y+y\dot y+\beta y^3 =0
\end{equation}

\noindent
where $\beta$ is a constant. For instance,
the problem of a causal viscous fluid with the bulk viscosity coefficient
$\zeta$ proportional to $\rho^{1/2}$ and $y\propto H$
in the truncated Extended Irreversible Thermodynamics theory \cite{visco}.
Also, the behavior near the singularity, when the
relaxation term is much more important than the viscous term in the transport
equation, for generic power-law
relation $\zeta=\alpha\rho^m$ \cite{Zak}.
For a time decaying cosmological "constant",
$\dot\Lambda\sim -H^3\Lambda$ \cite{Reut}.
In a phenomenological description of the reheating process in
terms of an out-of-equilibrium mixture of two reacting fluids
\cite{ZPM}.

Equation (\ref{1}) is also very interesting from the mathematical point of
view as it appears in the analysis of the Painlev\'e equations \cite{Ince},
and it is the simplest case of a class of nonlinear differential equations
that possess form invariance under nonlocal transformations, so that they can
be linearized and its general solution can be obtained in parametric form
\cite{Ale} \cite{Chi97a}.

The general solution of equation (\ref{1}) is

\begin{equation} \label{yeta}
y(\eta)=\left(A e^{\lambda_+\eta}+Be^{\lambda_-\eta}\right)^{1/2}
\end{equation}
\begin{equation} \label{teta}
\Delta t(\eta)=\int \frac{d\eta}{y(\eta)}
\end{equation}

\noindent where
$\lambda_{\pm}=\left(1/2\right)\left[-1\pm\left(1-8\beta\right)^{1/2}\right]$
are  the  roots  of  the  characteristic  polynomial of the linear equation
and $A$ and $B$ are two arbitrary integration constants. The analysis of
the general solution shows that it possesses, for generic $\beta$, movable
singularities with asymptotic behavior

\begin{equation} \label{ysing}
y(t)\sim \frac{\alpha}{\Delta t} \sum_{n=0}^\infty c_n(\beta) \gamma^n
\Delta t^{nr}
\end{equation}

\noindent where $r=4-\alpha>0$ is the Kowalevski exponent (\cite{Yosh}),
$\alpha=\alpha_-$ for $\eta\to -\infty$ and $\alpha=\alpha_+$ for $\eta\to
\infty$ ($\beta<0$) and $\alpha_\pm=-2/\lambda_\pm$, $c_0=1$ and $\gamma$ is
an arbitrary integration constant. Thus (\ref{1}) is a second order
differential equation of the form (\ref{dy}) for which we can show that
2-parameter families of solutions exist with an expansion like (\ref{yt}).
Then it is interesting to see how the algorithm described above works in this
case.

We begin by writing equation (\ref{1}) in the Maple worksheet

\begin{verbatim}
> d:=diff(y(t),t$2)+y(t)*diff(y(t),t)+beta*y(t)^3;
\end{verbatim}
\[
d := {\frac {\partial ^{2}}{\partial t^{2}}}\,{\rm y}(t) +
{\rm y}(t)\,{\frac {\partial }{\partial t}}\,{\rm y}(t) + \beta
 \,{\rm y}(t)^{3}
\]

\noindent
and then we begin the iteration.

\subsection{Leading term}

For $M=1$, inserting the leading term in (\ref{1}), we get three exponents

\begin{verbatim}
> f:=collect2(subs(y(t)=c1*t^n1,d),t,exponents);
\end{verbatim}
\[
f := [n1 - 2, \,2\,n1 - 1, \,3\,n1]
\]

\noindent
We look for the values of $n1$ such that two or more terms balance

\begin{verbatim}
> n1e:=balance(f,n1);
\end{verbatim}
\[
{\it n1e} := [-1]
\]

\noindent
This shows that all three terms balance simultaneously for $n1=-1$.

\begin{verbatim}
> f:=collect2(subs(y(t)=c1*t^(-1),d),t,exponents);
\end{verbatim}
\[
f := [-3]
\]

\noindent
Then $e_1=-3$, and $c1$ satisfies a quadratic

\begin{verbatim}
> _coeff[-3];
> c1s:={solve(",c1)}minus{0};
\end{verbatim}
\[
2\,{\it c1} - {\it c1}^{2} + \beta \,{\it c1}^{3}
\]
\[
{\it c1s} := \{{\displaystyle \frac {1}{2}} \,{\displaystyle 
\frac {1 + \sqrt{1 - 8\,\beta }}{\beta }} , \,{\displaystyle 
\frac {1}{2}} \,{\displaystyle \frac {1 - \sqrt{1 - 8\,\beta }}{
\beta }} \}
\]

\noindent
Thus, two families of real solutions with this leading behavior exist provided
$\beta<1/8$.

\subsection{Second term}

For $M=2$, using $c_1, n_1$ from the previous step, we look for the exponent
following $-3$.

\begin{verbatim}
> f:=collect2(subs(y(t)=c1/t+c2*t^n2,d),t,exponents);
\end{verbatim}
\[
f := [-3, \,n2 - 2, \,2\,n2 - 1, \,3\,n2]
\]

\noindent
These exponents are linear functions of $n2$, and become equal only for
$n2=-1$. As we require that $n2>n1=-1$, we can sort them

\begin{verbatim}
> assume(n2> -1):
> sort(f,(a,b)-> is(a<b));
\end{verbatim}
\[
[-3, \,n2 - 2, \,2\,n2 - 1, \,3\,n2]
\]

\noindent
and find that $e_2=n_2-2$.
Its coefficient

\begin{verbatim}
> _coeff[n2-2];
\end{verbatim}
\[
{\it c2}\,n2^{2} - {\it c2}\,n2 + {\it c2}\,{\it c1}\,n2 - {\it c2}
\,{\it c1} + 3\,\beta \,{\it c1}^{2}\,{\it c2}
\]

\noindent
is linear in $c2$, so that this parameter remains free. Then we find
$n2=3-c1$, where the root $n2=-2$ is discarded as it arises from time
translation invariance of one-parameter solutions. On the other hand, as $ - 1
< n2$, this implies ${\it c1} < 4$, so that $c1_-$ is admissible for
$\beta<1/8$, with $-1<n_2< 3$; while $c1_+$ is admissible for $\beta<0$, with
$n_2>3$.

\subsection{Third term}

For $M=3$, we look for the third exponent

\begin{verbatim}
> f:=collect2(subs(y(t)=c1/t+c2*t^n2+c3*t^n3,d),t,exponents);
\end{verbatim}
\begin{eqnarray*}
\lefteqn{f := [-3, \,{\it n2} - 2, \,{\it n3} - 2, \,2\,{\it n2} - 1, \,2
\,{\it n3} - 1, \,2\,{\it n2} + {\it n3}, \,{\it n2} + 2\,{\it n3
}, } \\
 & & {\it n2} + {\it n3} - 1, \,
3\,{\it n2}, \,3\,{\it n3}] \mbox{\hspace{240pt}}
\end{eqnarray*}

\noindent These are linear functions of $n3$, where we must take into account
that $n2<n3$ and $-1<n_2(\beta)<\infty$. Let us see first when these exponents
become equal.

\begin{verbatim}
> n3e:=balance(f,n3);
\end{verbatim}
\begin{eqnarray*}
\lefteqn{{\it n3e} := [-1, \,{\it n2}, \, - {\displaystyle
\frac {3}{2}}  - {\displaystyle \frac {1}{2}} \,{\it n2}, \, - 3
 - 2\,{\it n2}, \,1 + 2\,{\it n2}, \,{\displaystyle \frac {1}{2}
} \,{\it n2} - {\displaystyle \frac {1}{2}} , \, - 2 - {\it n2}, 
} \\
 & & {\displaystyle \frac {1}{3}} \,{\it n2} - {\displaystyle
\frac {2}{3}} ,  
\,2 + 3\,{\it n2},
{\displaystyle \frac {2}{3}} \,{\it n2} - {\displaystyle 
\frac {1}{3}} , \,{\displaystyle \frac {1}{2}}  + {\displaystyle 
\frac {3}{2}} \,{\it n2}]\mbox{\hspace{150pt}}
\end{eqnarray*}

\noindent
These in turn  become equal when $n2= - 1$.
\begin{verbatim}
n2e:=balance(n3e,n2);
\end{verbatim}
\[
{\it n2e} := [-1]
\]

\noindent
Hence we can sort them

\begin{verbatim}
> assume(n2>-1):
> sort(n3e,(a,b)->is(a<b));
\end{verbatim}
\begin{eqnarray*}
\lefteqn{[ - 3 - 2\,{\it n2}, \, - 2 - {\it n2}, \, - 
{\displaystyle \frac {3}{2}}  - {\displaystyle \frac {1}{2}} \,
{\it n2}, \,-1, \,{\displaystyle \frac {1}{3}} \,{\it n2} - 
{\displaystyle \frac {2}{3}} , \, - {\displaystyle \frac {1}{2}} 
 + {\displaystyle \frac {1}{2}} \,{\it n2},   } \\
 & & {\displaystyle
\frac {2}{3}} \,{\it n2} - {\displaystyle \frac {1}{3}} , \,{\it 
n2}, \,
{\displaystyle \frac {1}{2}}  + {\displaystyle \frac {3}{2
}} \,{\it n2}, \,2\,{\it n2} + 1, \,
3\,{\it n2} + 2]\mbox{\hspace{200pt}}
\end{eqnarray*}

\noindent
We look for the third exponent within each interval.

\begin{verbatim}
> l:=[ n2, 1/2+3/2*n2, 2*n2+1,3*n2+2,3*n2+2+1]:
> l1:=subs(n2=.3,l):
for i from 1 to nops(l1)-1 do
u:=(l1[i]+l1[i+1])/2;
tabla[i]:=[[l[i]<n3,n3<l[i+1]],
op(3,sort(f,(a,b)->ord2(a,b,n2=.3,n3=u)))]:
print(tabla[i]);
od:
\end{verbatim}
\[
[[{\it n2} < {\it n3}, \,{\it n3} < {\displaystyle \frac {1}{2}} 
 + {\displaystyle \frac {3}{2}} \,{\it n2}], \,{\it n3} - 2]
\]
\[
[[{\displaystyle \frac {1}{2}}  + {\displaystyle \frac {3}{2}} \,
{\it n2} < {\it n3}, \,{\it n3} < 2\,{\it n2} + 1], \,{\it n3} - 
2]
\]
\[
[[2\,{\it n2} + 1 < {\it n3}, \,{\it n3} < 3\,{\it n2} + 2], \,2
\,{\it n2} - 1]
\]
\[
[[3\,{\it n2} + 2 < {\it n3}, \,{\it n3} < 3\,{\it n2} + 3], \,2
\,{\it n2} - 1]
\]

\noindent
and we find that there is balance when
${\it n3}=2\,{\it n2} + 1$ with coefficient

\begin{verbatim}
> expand(subs(n3=2*n2+1,n2=3-c1,_coeff[n3-2]+_coeff[2*n2-1])):
> simplify(",{2-c1+beta*c1^2});
\end{verbatim}
\[
3\,\beta \,{\it c1}\,{\it c2}^{2} + 36\,{\it c3} + 3\,{\it c2}^{2
} + ( - 17\,{\it c3} - {\it c2}^{2})\,{\it c1} + 2\,{\it c3}\,
{\it c1}^{2}
\]

\noindent
Then $ - 1 <  n3(\beta) < \infty$ and
${\it n3}=2\,({\it n2} + 1) - 1$ so that we recover the first three terms of
(\ref{ysing}) with

\[c3=
 - {\displaystyle \frac {{\it c2}^{2}\,(3\,\beta \,{\it c1} + 3
 - {\it c1})}{36 - 17\,{\it c1} + 2\,{\it c1}^{2}}} 
\]

\noindent proportional to $c_2^2$ as expected from (\ref{ysing}). Also, as
expected, no other solution is found.

\section{Equation with a variable exponent}

In order to treat dissipative processes in cosmology which are not close to
equilibrium a nonlinear phenomenological generalization of the Israel-Stewart
theory was developed recently \cite{mm}. Processes for which this kind of
processes may have occured are inflation driven by a viscous stress
\cite{mm,m}, and the reheating era at the
end of inflation \cite{ZPM}.

In a spacially flat FLRW universe, Einstein's equations together with state
and transport equations of the fluid
give the evolution equation for the Hubble rate \cite{mm}

\begin{eqnarray}\label{14}
&& \left[1-{k^2\over v^2}-\left({2k^2\over3\gamma v^2}\right){\dot{H}
\over H^2}\right]\left\{\ddot{H}+3H\dot{H}+\left(
{1-2\gamma\over\gamma}\right){\dot{H}^2\over H}+{9\over4}
\gamma H^3\right\}  \nonumber\\
&&{}+{3\gamma v^2\over2\alpha}\left[1+\left(
{\alpha k^2\over\gamma v^2}
\right)H^{q-1}\right]H^{2-q}\left(2\dot{H}+3\gamma H^2\right)
-{9\over2}\gamma v^2H^3=0
\end{eqnarray}

\noindent where $\gamma$, $\alpha$, $v$, $q$ and $k$ are parameters describing
the thermodynamical properties of the fluid. Note that $q$ appears in the
exponent.

Let us insert equation (\ref{14}) in a new worksheet.

\begin{verbatim}
> d:=(1-k^2/v^2-2*k^2/(3*gamma*v^2)*diff(H(t),t)/H(t)^2)*
(diff(H(t),t$2)+3*H(t)*diff(H(t),t)+
(1-2*gamma)/gamma*diff(H(t),t)^2/H(t)+9/4*gamma*H(t)^3)+
(3*gamma*v^2/(2*alpha)*(1+alpha*k^2/(gamma*v^2)*
H(t)^(q-1))*H(t)^(2-q)*(2*diff(H(t),t)+
3*gamma*H(t)^2)-9/2*gamma*v^2*H(t)^3):
\end{verbatim}

\noindent Multiplication by a factor $H^3$ makes all powers of
$H$ positive.

\begin{verbatim}
> d2:=map(simplify,expand1(H(t)^3*d),power,symbolic):
> dview(d2,t,H);
\end{verbatim}
\begin{eqnarray*}
\lefteqn{H^{3}\,H^{^{ {\prime \prime }}} + 3\,H^{4}\,H^{
{}^{\prime }} + {\displaystyle \frac {H^{2}\,(H^{{}^{\prime }})^{2}}{
\gamma }}  - 2\,H^{2}\,(H^{^{\prime }})^{2} + {\displaystyle
\frac {9}{4}} \,H^{6}\,\gamma  - {\displaystyle \frac {H^{3}\,k^{
2}\,H^{{^{\prime \prime }}}}{v^{2}}} } \\
 & & \mbox{}  - {\displaystyle
\frac {9}{2}} \,{\displaystyle \frac {H^{4}\,k^{2}\,H^{{}^{\prime }
}}{v^{2}}} 
- 3\,{\displaystyle \frac {H^{2}\,k^{2}\,(H^{{}
^{\prime }})^{2}}{\gamma \,v^{2}}}  + 2\,{\displaystyle \frac {H
^{2}\,k^{2}\,(H^{^{\prime }})^{2}}{v^{2}}}  - {\displaystyle
\frac {9}{4}} \,{\displaystyle \frac {H^{6}\,k^{2}\,\gamma }{v^{2
}}}  \\
 & & \mbox{}
 - {\displaystyle \frac {2}{3}} \,{\displaystyle \frac {H\,k
^{2}\,H^{^{\prime }}\,H^{{\it ^{\prime \prime }}}}{\gamma \,v
^{2}}}
- {\displaystyle \frac {2}{3}} \,{\displaystyle \frac {k
^{2}\,(H^{^{\prime }})^{3}}{\gamma ^{2}\,v^{2}}}  + 
{\displaystyle \frac {4}{3}} \,{\displaystyle \frac {k^{2}\,(H^{
^{\prime }})^{3}}{\gamma \,v^{2}}} + 3\,{\displaystyle \frac {H^{(5 - q)}\,\gamma \,v^{
2}\,H^{^{\prime }}}{\alpha }} \\
 & & \mbox{}  + {\displaystyle \frac {9}{2}} \,
{\displaystyle \frac {H^{(7 - q)}\,\gamma ^{2}\,v^{2}}{\alpha }} 
 + 3\,H^{4}\,k^{2}\,H^{^{\prime }} + {\displaystyle \frac {9}{2}
} \,H^{6}\,\gamma \,k^{2} - {\displaystyle \frac {9}{2}} \,H^{6}
\,\gamma \,v^{2}\mbox{\hspace{90pt}}
\end{eqnarray*}

\subsection{Leading term}

\begin{verbatim}
> f:=collect2(subs(H(t)=c1*t^n1,d2),t,exponents,
x->collect(x,[k,c1,v]));
\end{verbatim}

\[
f := [6\,{\it n1}, \,4\,{\it n1} - 2, \,5\,{\it n1} - 1, \,3\,
{\it n1} - 3, \, - {\it n1}\,( - 7 + q), \,6\,{\it n1} - {\it n1}
\,q - 1]
\]

\noindent
The values of $n1$ that make these terms balance are

\begin{verbatim}
> n1e:=balance(f,n1);
\end{verbatim}

\[
{\it n1e} := [0, \,-1, \,{\displaystyle \frac {2}{ - 3 + q}} , \,
{\displaystyle \frac {1}{ - 2 + q}} , \,{\displaystyle \frac {3}{
 - 4 + q}} , \, - {\displaystyle \frac {1}{q}} ]
\]

\noindent
These in turn are function of $q$, so we look whether there is a critical
value of $q$

\begin{verbatim}
> qe:=balance(n1e,q);
\end{verbatim}

\[
{\it qe} := [1]
\]

\noindent This shows that $q=1$ delimits different behaviors of the solutions
as it was already known from investigation of equation
(\ref{14}) \cite{cjmm}. To be concise, in the following only the case $q<1$
will be considered. Further, sorting must be done separately for each interval
$(-\infty,0), (0,1), (1,2), (2,3), (3,4), (4,\infty)$, so we will restrict to
the case $0<q<1$.

\begin{verbatim}
> n1ea:=sort(n1e,(a,b)->ord1(a,b,q=.5));
\end{verbatim}
\[
{\it n1ea} := [ - {\displaystyle \frac {1}{q}} , \,-1, \,
{\displaystyle \frac {3}{ - 4 + q}} , \,{\displaystyle \frac {2}{
 - 3 + q}} , \,{\displaystyle \frac {1}{ - 2 + q}} , \,0]
\]

\noindent
Now we look for the leading exponent within each interval.

\begin{verbatim}
l:=[n1ea[1]-1,op(n1ea),n1ea[-1]+1]:
l1:=subs(q=.5,l):
for i from 1 to nops(l1)-1 do
u:=(l1[i]+l1[i+1])/2;
tabla[i]:=[[l[i]<n1,n1<l[i+1]],
op(1,sort(f,(a,b)->ord2(a,b,q=.5,n1=u)))]:
print(tabla[i]);
od:
\end{verbatim}
\[
[[ - {\displaystyle \frac {1}{q}}  - 1 < {\it n1}, \,{\it n1} < 
 - {\displaystyle \frac {1}{q}} ], \, - {\it n1}\,( - 7 + q)]
\]
\[
[[ - {\displaystyle \frac {1}{q}}  < {\it n1}, \,{\it n1} < -1], 
\, - {\it n1}\,( - 7 + q)]
\]
\[
[[-1 < {\it n1}, \,{\it n1} < {\displaystyle \frac {3}{ - 4 + q}
} ], \,6\,{\it n1} - {\it n1}\,q - 1]
\]
\[
[[{\displaystyle \frac {3}{ - 4 + q}}  < {\it n1}, \,{\it n1} < 
{\displaystyle \frac {2}{ - 3 + q}} ], \,6\,{\it n1} - {\it n1}\,
q - 1]
\]
\[
[[{\displaystyle \frac {2}{ - 3 + q}}  < {\it n1}, \,{\it n1} < 
{\displaystyle \frac {1}{ - 2 + q}} ], \,3\,{\it n1} - 3]
\]
\[
[[{\displaystyle \frac {1}{ - 2 + q}}  < {\it n1}, \,{\it n1} < 0
], \,3\,{\it n1} - 3]
\]
\[
[[0 < {\it n1}, \,{\it n1} < 1], \,3\,{\it n1} - 3]
\]

\noindent
Thus we find that the leading exponent switch at $n1=-1$ and $n1=2/(-3+q)$,
where terms  balance. Let us start with $n1=-1$.

\begin{verbatim}
> f:=collect2(subs(H(t)=c1*t^(-1),d2),t,exponents,
x->collect(x,[k,c1,v])):
> assume(q<1):
> sort(f,(a,b)->is(a<b));
\end{verbatim}
\[
[ - 7 + {\it q}, \,-6]
\]

\begin{verbatim}
> _coeff[-7+q];
\end{verbatim}
\[
( - 3\,{\displaystyle \frac {{\it c1}^{(6 - q)}\,\gamma }{\alpha 
}}  + {\displaystyle \frac {9}{2}} \,{\displaystyle \frac {{\it 
c1}^{(7 - q)}\,\gamma ^{2}}{\alpha }} )\,v^{2}
\]

\noindent
Thus, $n_1=-1$ and $c_1=2/(3\gamma)$ gives a leading behavior.
Let us see the other case.

\begin{verbatim}
> f:=collect2(subs(H(t)=c1*t^(2/(-3+q)),d2),t,exponents,
x->collect(x,[k,c1,v])):
> sort(f,(a,b)->ord1(a,b,q=.5));
\end{verbatim}
\[
[ - 3\,{\displaystyle \frac { - 5 + q}{ - 3 + q}} , \, - 2\,
{\displaystyle \frac { - 7 + q}{ - 3 + q}} , \, - {\displaystyle 
\frac { - 13 + q}{ - 3 + q}} , \,{\displaystyle \frac {12}{ - 3
 + q}} ]
\]

\begin{verbatim}
> _coeff[-3*(-5+q)/(-3+q)];
\end{verbatim}
\[
 6\,{\displaystyle \frac {{\it c1}^{(6 - q)}\,\gamma
\,v^{2}}{\alpha \,( - 3 + q)}}  + {\displaystyle \frac {(
{\displaystyle \frac {16}{3}} \,{\displaystyle \frac {1}{\gamma 
\,( - 3 + q)^{3}}}  + {\displaystyle \frac {8}{3}} \,
{\displaystyle \frac {1}{\gamma \,( - 3 + q)^{2}}}  - 
{\displaystyle \frac {16}{3}} \,{\displaystyle \frac {1}{\gamma 
^{2}\,( - 3 + q)^{3}}} )\,{\it c1}^{3}\,k^{2}}{v^{2}}} 
\]

\noindent
Thus $n_1=2/(-3+q)$ and

\[
c1=( - {\displaystyle \frac {9}{4}} \,{\displaystyle \frac {\gamma
^{3}\,{\it v}^{4}\,(9 - 6\,q + q^{2})}{{\it k
}^{2}\,\alpha \,( - \gamma  + \gamma \,q
 - 2)}} )^{(\frac {1}{ - 3 + q})}
\]

\noindent give the leading behavior of another family of solutions
that, for the sake of brevity, we will not pursue further in the step $M=2$.
No solution is found when $n1$ is between the balancing values.

\subsection{Second term for $n_1=-1$}

\noindent
We start by expanding the terms with non integer exponents.

\begin{verbatim}
> d3:=subs({H(t)=c1/t+c2*t^n2,
H(t)^(5-q)=(c1/t)^(5-q)*(1+c2/c1*(5-q)*t^(n2+1)),
H(t)^(7-q)=(c1/t)^(7-q)*(1+c2/c1*(7-q)*t^(n2+1))},d2):
\end{verbatim}

\noindent
Then the exponents are

\begin{verbatim}
> f:=collect2(d3,t,exponents,x->collect(x,[k,c1,c2,v,n2]));
\end{verbatim}
\begin{eqnarray*}
\lefteqn{f := [-6, \, - 2 + 4\,{\it n2}, \, - 4 + 2\,{\it n2}, \,
 - 7 + q, \, - 3 + 3\,{\it n2}, \,6\,{\it n2},  } \\
 & &  \, - 5 + q + 2\,{\it n2},
\, - 1 + 5\,{\it n2}, \, - 5 + {\it n2},
 - 6 + q + {\it n2}]\mbox{\hspace{90pt}}
\end{eqnarray*}

\noindent
These are the balancing values of $n2$

\begin{verbatim}
> n2e:=balance(f,n2):
> assume(q<1):
> sort(n2e,(a,b)->is(a<b));
\end{verbatim}
\[
[ - 2 + q, \, - {\displaystyle \frac {3}{2}}  + {\displaystyle 
\frac {q}{2}}, \, - {\displaystyle \frac {4}{3}}  +
{\displaystyle \frac {q}{3}}, \, - {\displaystyle \frac {5}{4
}}  + {\displaystyle \frac {q}{4}}, \, - {\displaystyle
\frac {6}{5}}  + {\displaystyle \frac {q}{5}}, \, -
{\displaystyle \frac {7}{6}}  + {\displaystyle \frac {q}{6}},
\,-1, \, - {\displaystyle \frac {1}{2}}  - {\displaystyle
\frac {q}{2}}, \, - q]
\]

\noindent As we require $n2>-1$, only two balancing values are admissible
$n2=-q$ and $ n2= - \frac {1}{2} - \frac {q}{2}$.
In the case $n2=-q$

\begin{verbatim}
> f:=collect2(subs(n2=-q,d3),t,exponents,
x->collect(x,[k,c1,c2,v,q])):
> assume(q<1):
> sort(f,(a,b)->is(a<b));
\end{verbatim}
\[
[ - 7 + q, \,-6, \, - 5 - q, \, - 4 - 2\,q, \, - 3 - 3\,q, \, - 2
 - 4\,q, \, - 1 - 5\,q, \, - 6\,q]
\]

\noindent
it turns out that $e_2=-6$. From its coefficient

\begin{verbatim}
> _coeff[-6]:
expand("/c1^3):
simplify(subs(c1=2/(3*gamma),"));
\end{verbatim}
\[
 - {\displaystyle \frac {4}{3}} \,{\displaystyle \frac {v^{2}\,(
\alpha  + 3^{q}\,\gamma ^{(q + 1)}\,{\it c2}\,2^{( - q)}\,q - 6\,
3^{(q - 1)}\,\gamma ^{(q + 1)}\,{\it c2}\,2^{( - q)})}{\gamma ^{2
}\,\alpha }} 
\]

\noindent
we find $c2=
\frac {\alpha }{\gamma \,(2 - q)}(\frac {2}{3\,\gamma })^{q}$.
In the case $n2=-\frac {1}{2} - \frac {q}{2}$

\begin{verbatim}
> f:=collect2(subs(n=1/2+q/2,d3),t,exponents,
x->collect(x,[k,c1,c2,v,q])):
> assume(q<1):
> sort(f,(a,b)->is(a<b));
\end{verbatim}
\begin{eqnarray*}
\lefteqn{[ - 7 + q, \, - {\displaystyle \frac {13}{2}}  + {\displaystyle
\frac {q}{2}},\,-6, \, - {\displaystyle \frac {11}{2}}  -
{\displaystyle \frac {q}{2}}, \,- 5 - q, \, -
{\displaystyle \frac {9}{2}}  - {\displaystyle \frac {3}{2}} \,q
, \, - 4 - 2\,q,} \\
 & &  - {\displaystyle \frac {7}{2}}  -
{\displaystyle \frac {5}{2}} \,q, \, - 3 - 3\,q]
   \mbox{\hspace{200pt}}
\end{eqnarray*}

\noindent
we find $e_2=-13/2+q/2$. Its coefficient however only vanishes for
$c2=0$.

\subsection{Solutions}

We collect here the solutions of eq. (\ref{14}) found thus far, including
those for $q>1$ whose calculation has not been given in detail in this
section. We give first the solutions with $n_1=-1$.

\noindent
Case $2-q<n_2<1$:

\begin{equation} \label{4}
c_1=\frac{2}{3\gamma}\frac{1}{\sqrt{2}v+1}, \qquad
n_2=\frac{\sqrt{2}\left(\gamma-2\right)v+\gamma}
{\gamma\left(\sqrt{2}v+1\right)}
\end{equation}

\noindent In this case,
$\left(1/3\right)\left(1/\left(1+\sqrt{2}\right)\right)\le c_1<2/3$, and
$\left(1-\sqrt{2}\right)/\left(1+\sqrt{2}\right)\le n<1$. We are assuming that
$c_2\neq 0$.

\bigskip
\noindent
Case $n_2=q<1$:

\begin{equation} \label{5}
c_1=\frac{2}{3\gamma},\qquad
c_2=\frac{\alpha}{(2-q)\gamma }\left(\frac{2}{3\gamma}\right)^q
\end{equation}

\bigskip
\noindent
Case $n_2=2-q<1$:

\noindent
There are three subcases.

\noindent
a)
\begin{equation} \label{6}
c_1=\frac{2}{3\gamma}\frac{k^2}{k^2-v^2},\qquad
c_2=-\frac{8k^{6-2q}\left(k^2-v^2\right)^{q-3}v^4}
{9\gamma\alpha q\left(2k^2-v^2\right)}
\left(\frac{3\gamma}{2}\right)^q
\end{equation}

\noindent
with $k>v$.

\noindent
b)
\begin{equation} \label{7}
c_1=\frac{2}{3\gamma}\frac{1}{1+\sqrt{2}v}
\end{equation}

\noindent
and $c_2$ a long expression of the form $N/D$, with

$$N=8\sqrt{2}v^4\left(\sqrt{2}v-1\right)^2\left(3\gamma/2\right)^q
\left(\sqrt{2}v+1\right)^{q-3}$$

$$
D=9\gamma q\alpha\left\{2\sqrt{2}\left[\left(q-1\right)\gamma-2\right]v^4
+\left[\gamma\left(2k^2-1\right)\left(q-1\right)+4\left(1-k^2\right)
\right]2v^3
\right.
$$
$$
+\left[-\sqrt{2}\left(1+2k^2\right)\left(q-1\right)+
2\sqrt{2}\left(4k^2-1\right)\right]v^2
$$
$$
\left.+\left[\gamma\left(1-2k^2\right)\left(q-1\right)-4k^2\right]v+
\sqrt{2}\left(q-1\right)k^2\gamma\right\}
$$

\noindent
c)

\begin{equation} \label{8}
c_1=\frac{2}{3\gamma}\frac{1}{1-\sqrt{2}v}
\end{equation}

\noindent
with $v<\sqrt{2}$ and $c_2$ a very long expression that is omitted here.

On the other hand, for $n_1=2/(q-3)$ we get

$$
c_1= \left[\frac{9\gamma^3v^4\left(9-6q+q^2\right)}
{4k^2\alpha\left(2+\gamma-\gamma q\right)}\right]^{\frac{1}{q-3}}
$$

\noindent
where $q<1$.

\section{Anisotropic universe with a  scalar field}

Homogeneous anisotropic cosmological models with a self-interacting scalar
field has been investigated recently to verify the generality of inflationary
solutions. For this purpose it has been found convenient to cast the metric in
the semiconformal form \cite{AFI}. In these coordinates the Bianchi VI${}_0$
metric becomes

\begin{equation} \label{ds}
ds^2=e^{f(t)}\left( -dt^2+dz^2\right) + G(t)\left( e^zdx^2+e^{-z}dy^2\right)
\end{equation}

A scalar field with exponential potential $V=V_0 e^{k\phi}$ is interesting
from the physical point of view because it arises in several particle
theories, in the effective four-dimensional theories induced by Kaluza-Klein
theories, including various higher-dimensional supergravity \cite{salam} and
superstring \cite{fradkin} models \cite{olive,taha,easther}. It is also
interesting from the mathematical point of view because it allows decoupling
of the geometric and matter degrees of freedom \cite{Chi97b}. Thus the whole
evolution of the model is obtained from the solution of a third order ODE for
$G(t)$

\begin{equation} \label{dG}
\ddot G^2 G-K\ddot G\dot G^2-\tres G\dot G G+\frac{1}{2}\ddot G G^2+
m^2\ddot G=0;   
\end{equation}

\noindent
where $K=\frac{k^2}{4}-\frac{1}{2}$ and $m$ is an arbitrary integration
constant. We will investigate here the singular behavior of this model,
corresponding to $G(t)\to 0$ for $t\to t_0$. Then we write eq. (\ref{dG}) in a
new worksheet.

\begin{verbatim}
> d:=diff(G(t),t$2)^2*G(t)-K*diff(G(t),t$2)*diff(G(t),t)^2
-diff(G(t),t$3)*diff(G(t),t)*G(t)+1/2*diff(G(t),t$2)*G(t)^2
+m^2*diff(G(t),t$2):
> dview(d,t,G);
\end{verbatim}
\[
(G^{{}^{\it {\prime \prime }}})^{2}\,G - K\,G^{{}^{\it {\prime \prime }}}\,
(G^{{}^{\prime }})^{2} - G^{{}^{\it {\prime \prime
\prime }}}\,G^{{}^{\prime }}\,G + {\displaystyle \frac {1}{2}} \,
G^{{}^{\it {\prime \prime }}}\,G^{2} + m^{2}\,G^{{}^{\it {\prime \prime }}}
\]

\subsection{Leading term}

Inserting the leading term in (\ref{dG}) we obtain the exponents

\begin{verbatim}
> f:=collect2(subs(G(t)=c1*t^n1,d),t,exponents);
\end{verbatim}
\[
f := [3\,{\it n1} - 4, \,3\,{\it n1} - 2, \,{\it n1} - 2]
\]

\noindent
The values of $n1$ that make these terms balance are

\begin{verbatim}
> n1e:=balance(f,n1);
\end{verbatim}
\[
{\it n1e} := [0, \,1]
\]

\noindent
and the leading exponent inside each interval is

\begin{verbatim}
> l1:=[-infinity,op(n1e),infinity]:
for i from 1 to nops(n1e)+1 do assume(l1[i]<n1,n1<l1[i+1]);
tabla[i]:=[[l1[i]<n1,n1<l1[i+1]],op(1,sort(f,(a,b)->is(a<b)))]:
print(tabla[i]);
od:
\end{verbatim}
\[
[[ - \infty  < {\it n1}, \,{\it n1} < 0], \,3\,{\it n1} - 4]
\]
\[
[[0 < {\it n1}, \,{\it n1} < 1], \,3\,{\it n1} - 4]
\]
\[
[[1 < {\it n1}, \,{\it n1} < \infty ], \,{\it n1} - 2]
\]

\noindent
Thus the leading behavior changes at $n=1$. Let us consider first that
$n1<1$.

\begin{verbatim}
> _coeff[3*n1-4];
\end{verbatim}
\[
{\it c1}^{3}\,{\it n1}^{3} - {\it c1}^{3}\,{\it n1}^{2} - K\,
{\it c1}^{3}\,{\it n1}^{4} + K\,{\it c1}^{3}\,{\it n1}^{3}
\]

\noindent
We obtain $n1=1/K$ and $c1$ remains free. This implies $1/K<1$, that is
either $1/2<K<0$ or $K>1$. On the other hand, the coefficient for  $n1>1$

\begin{verbatim}
> _coeff[n1-2];
\end{verbatim}
\[
m^{2}\,{\it c1}\,{\it n1}^{2} - m^{2}\,{\it c1}\,{\it n1}
\]

\noindent shows that this case can occur only for $m=0$. In such a case $n1$
remains free.
We reject the balancing value $n1=0$ because it does not yield a singular
behavior. We note however that $G=c1$ is an exact solution. The other
balancing value $n1=1$ is quite special as $G=c_1 t$ is also an exact solution
with $c_1$ arbitrary.

\subsection{Second term for $n1=1$}

We obtain the exponents

\begin{verbatim}
> f:=collect2(subs(G(t)=c1*t+c2*t^n2,d),t,exponents);
> assume(1<n2):
> sort(f,(a,b)->is(a<b));
\end{verbatim}
\[
[{\it n2} - 2, \,2\,{\it n2} - 3, \,3\,{\it n2} - 4, \,{\it n2}, 
\,2\,{\it n2} - 1, \,3\,{\it n2} - 2]
\]

\noindent
and we find that $e_2=n_2-2$ for $n_2>1$, with a coefficient

\begin{verbatim}
> _coeff[n2-2];
\end{verbatim}
\begin{eqnarray*}
\lefteqn{m^{2}\,{\it c2}\,{\it n2}^{2} - m^{2}\,{\it c2}\,{\it n2
} - {\it c2}\,{\it n2}^{3}\,{\it c1}^{2} - K\,{\it c2}\,{\it n2}
^{2}\,{\it c1}^{2} + K\,{\it c2}\,{\it n2}\,{\it c1}^{2}
} \\
 & & \mbox{} - 2\,{\it c2}\,{\it n2}\,{\it c1}^{2}
+ 3\,{\it c2}\,{\it n2}^{2}\,{\it c1}^{2}
\mbox{\hspace{200pt}}
\end{eqnarray*}

\noindent Thus we find ${\it n2}=\frac {m^{2}}{{\it c1}^{2}} + 2 - K$ , $c2$
remains free and the constraint $\frac {m^{2}}{{\it c1}^{2}}+2-K>1$ must be
satisfied.

\subsection{Second term for $n1=1/K$}

In this case we obtain the exponents

\begin{verbatim}
> f:=collect2(subs(G(t)=c1*t^(1/K)+c2*t^n2,d),t,exponents);
\end{verbatim}
\begin{eqnarray*}
\lefteqn{f := [{\displaystyle \frac {2 - 2\,K + {\it n2}\,K}{K}} 
, \,{\displaystyle \frac {1 - 2\,K + 2\,{\it n2}\,K}{K}} , \,
{\it n2} - 2, \,{\displaystyle \frac {1 - 4\,K + 2\,{\it n2}\,K}{
K}} ,   } \\
 & & 3\,{\it n2} - 2, \, 3\,{\it n2} - 4, \,
{\displaystyle \frac {2 - 4\,K + {\it n2}\,K}{K}} , \, - 
{\displaystyle \frac { - 3 + 2\,K}{K}} , \, - {\displaystyle 
\frac { - 1 + 2\,K}{K}} ]\mbox{\hspace{45pt}}
\end{eqnarray*}

\noindent
The values of $n2$ that make them balance

\begin{verbatim}
> n2e:=balance(f,n2);
\end{verbatim}
\begin{eqnarray*}
\lefteqn{{\it n2e} := [{\displaystyle \frac {1}{3}} \,
{\displaystyle \frac {1}{K}} , \, - {\displaystyle \frac {K - 1}{
K}} , \,0, \,1, \,{\displaystyle \frac {1}{3}} \,{\displaystyle 
\frac {1 + 2\,K}{K}} , \,{\displaystyle \frac {1}{3}} \,
{\displaystyle \frac {2\,K + 3}{K}} , \,{\displaystyle \frac { - 
1 + 2\,K}{K}} , \,{\displaystyle \frac {1}{K}} , 
 } \\
 & & {\displaystyle \frac {3}{K}} , \, - {\displaystyle \frac { - 1 +
2\,K}{K}} , \, - {\displaystyle \frac {1}{K}} , \,
{\displaystyle \frac {1 + 2\,K}{K}} , \,{\displaystyle 
\frac {1 + K}{K}} ]\mbox{\hspace{160pt}}
\end{eqnarray*}

\noindent
depend on $K$, and in turn the values of $K$ that make them balance are

\begin{verbatim}
> Ke:=balance(n2e,K):
> sort(Ke,(a,b)->is(a<b));
\end{verbatim}
\[
[-3, \,-2, \,{\displaystyle \frac {-3}{2}} , \,-1, \,
{\displaystyle \frac {-2}{3}} , \,{\displaystyle \frac {-1}{2}} 
, \,{\displaystyle \frac {-1}{3}} , \,{\displaystyle \frac {1}{4}
} , \,{\displaystyle \frac {1}{3}} , \,{\displaystyle \frac {2}{5
}} , \,{\displaystyle \frac {1}{2}} , \,{\displaystyle \frac {2}{
3}} , \,1, \,{\displaystyle \frac {3}{2}} , \,2, \,3, \,4]
\]

\noindent Of these, only the values in $[-1/2,0)$ or $(1,\infty)$ are allowed.
Now we could sort the balancing values of $n2$ within each allowed interval of
$K$ where  it must be taken into account that $n2>1/K$. Clearly this analysis
is quite branched and we will not pursue it here. We just
consider the case of the balancing value $n2=1$.

\begin{verbatim}
> f:=collect2(subs(G(t)=c1*t^(1/K)+c2*t,d),t,exponents);
\end{verbatim}
\[
{\it f} := [ - {\displaystyle \frac { - 2 + K}{K}} , \,
{\displaystyle \frac {1}{K}} , \, - {\displaystyle \frac { - 3 + 
2\,K}{K}} , \, - {\displaystyle \frac { - 1 + 2\,K}{K}} ]
\]

\noindent
In this case the exponents depend on $K$ with balancing values

\begin{verbatim}
> balance(f,K);
\end{verbatim}

\[
[-1, \,1]
\]

\noindent
Now we sort the exponents within each allowed interval

\begin{verbatim}
> assume(-1/2<K,K<0):
> sort(f,(a,b)->is(a<b))
\end{verbatim}
\[
[ - {\displaystyle \frac { - 3 + 2\,K}{K}} , \, - {\displaystyle 
\frac { - 2 + K}{K}} , \, - {\displaystyle \frac { - 1 + 2\,K}{K}
} , \,{\displaystyle \frac {1}{K}} ]
\]

\begin{verbatim}
> assume(K>1):
> sort(f,(a,b)->is(a<b));
\end{verbatim}
\[
[ - {\displaystyle \frac { - 1 + 2\,K}{K}} , \, - {\displaystyle 
\frac { - 3 + 2\,K}{K}} , \, - {\displaystyle \frac { - 2 + K}{K}
} , \,{\displaystyle \frac {1}{K}} ]
\]

\noindent
For $K>1$ the coefficient is

\begin{verbatim}
> _coeff[-(-1+2*K)/K];
\end{verbatim}
\[
 - {\displaystyle \frac {{\it c2}^{2}\,{\it c1}}{K^{3}}}  + 
{\displaystyle \frac {m^{2}\,{\it c1}}{K^{2}}}  - {\displaystyle 
\frac {m^{2}\,{\it c1}}{K}}  + 3\,{\displaystyle \frac {{\it c1}
\,{\it c2}^{2}}{K^{2}}}  - 3\,{\displaystyle \frac {{\it c1}\,
{\it c2}^{2}}{K}}  + {\it c2}^{2}\,{\it c1}
\]

\noindent
and we get a solution: $n2=1$ and
$c2 =\pm\frac {\sqrt{K}\,m}{K - 1}$.

\section{Conclusions}

We have shown that approximations of solutions of nonlinear ordinary
differential equations relevant to cosmology can be obtained with the help of
CAS in the form of generalized power asymptotic expansions, without much
effort. For this purpose we have sketched an algorithm to make these
calculations by an iterative process.

The implementation of these calculations were made in Maple V Release 4. This
work environment is quite productive as it allows interactive exploration as
well as a powerful programming language. The set already written of procedures
to collect terms, obtain exponents and coefficients work in satisfactory way.
The next step that remains to be coded is sorting of exponents. This task is
more complex as it involves branching of options and has the further
difficulty that arises in the weakness of the current version of the {\bf
assume} facility.

From a more theoretical point of view, it deserves to be investigated how the
complexity of the algorithm increases with the order of iteration, and whether
this growth puts an effective limit to practical calculations. In such a case
it would be interesting to know whether more efficient algorithms can be
devised.

Also it would be interesting to know whether expansions of
solutions as shown in this paper, can give information about the integrability
of the equation.

\end{document}